\documentclass{amsart}
\usepackage{amsthm}
\usepackage{amsmath}
\usepackage{amsfonts}
\usepackage{amssymb}
\usepackage{mathrsfs}
\usepackage{hyperref}
\usepackage{fancyhdr}
\usepackage{graphicx}
\usepackage{float}                  
\usepackage{rotating}               
\restylefloat{figure}               
\theoremstyle{plain}

\theoremstyle{definition}

\def\N{{\mathbb N}}
\def\R{{\mathbb R}}

\def\Q{{\mathbb Q}}

\def\Z{{\mathbb Z}}

\def\T{{\mathbb T}}

\def\NN{\mathcal N}
\def\MM{{\mathcal M}}
\def\FF{{\mathcal F}}

\def\eps{\varepsilon}
\def\la{\lambda}
\def\Cci#1{C_c^\infty(#1)}

\def\lnorm#1{\left|\!\left|{#1}\right|\!\right|}
\def\dx{\,\text{d}x}
\def\dy{\,\text{d}y}
\def\d{\,\text{\rm d}}
\def\sigmaess{\sigma_{\text{\rm ess}}}

\def\supp{\text{\rm supp }}
\def\set#1#2{\left\{#1 \: ; \: #2\right\}}

\def\norm#1{\left|\!\left|{#1}\right|\!\right|}

\def\dist{\text{\rm dist}}

\def\phi{\varphi}
\def\theta{\vartheta}
\def\scapro#1#2{\left<#1,#2\right>}
\renewcommand\endproof{\hfill$\square$}
\begin{document}
\title{Spectral Properties of Grain Boundaries at Small Angles of Rotation}
\author{Rainer Hempel}
\author{Martin Kohlmann}
\address{Institute for Computational Mathematics, Technische
Universit\"at Braun\-schweig, Pockelsstra{\ss}e 14, 38106
Braunschweig, Germany} \email{r.hempel@tu-bs.de}
\address{Institute for Applied Mathematics, Leibniz Universit\"at Hannover, Welfengarten 1, 30167 Hannover, Germany}
\email{kohlmann@ifam.uni-hannover.de}
\begin{abstract} We study some spectral properties of a simple two-dimensional model for small angle defects
in crystals and alloys. Starting from a periodic potential $V
\colon \R^2 \to \R$, we let $V_\theta(x,y) = V(x,y)$ in the right
half-plane $\{ x \ge 0\}$
 and  $V_\theta = V \circ M_{-\theta}$ in the left half-plane $\{x < 0\}$,
 where $M_\theta \in \R^{2 \times 2}$ is the usual
matrix describing rotation of the coordinates in $\R^2$ by an
angle $\theta$. As a main result, it is shown that spectral gaps
of the periodic Schr\"odinger operator $H = -\Delta + V$ fill
with spectrum of $R_\theta = -\Delta + V_\theta$ as $0 \ne \theta
\to 0$.  Moreover, we obtain upper and lower bounds for a
quantity pertaining to an integrated density of states measure for
the surface states.
\end{abstract}
\maketitle
\tableofcontents
\section*{Introduction}
In the quantum theory of solids one first studies periodic
structures which can often be modelled by Schr\"odinger operators
with periodic potentials. Other models deal with situations where
periodicity holds only in subsets of the sample; more precisely,
the sample is the disjoint union of subsets such that, in each
 subset, the potential is obtained by restricting different periodic potentials
to the corresponding subsets. Such zones or ``grains'' occur
frequently in crystals and in alloys; some typical examples are
shown in Figure 1. It is an important issue to understand how the
interface between two grains will influence the energy spectrum of
the sample. Typically, the grain boundaries appear to be
(piecewise) linear, and one is led to study problems on $\R^2$
with a potential $W = W(x,y)$ defined by
$$
W(x,y) := \left\{%
\begin{array}{lll}
  V_r(x,y),    && \text{$x \ge 0$,} \\
  V_\ell(x,y), && \text{$x < 0$,} \\
\end{array}%
\right.\eqno{(0.1)}
$$
where $V_r, V_\ell \colon \R^2 \to \R$ are periodic. In many
situations, $V_\ell$ is obtained from $V_r$ by a translation or a
rotation about the origin.
\vskip5ex
\centerline{\includegraphics[width=5cm]{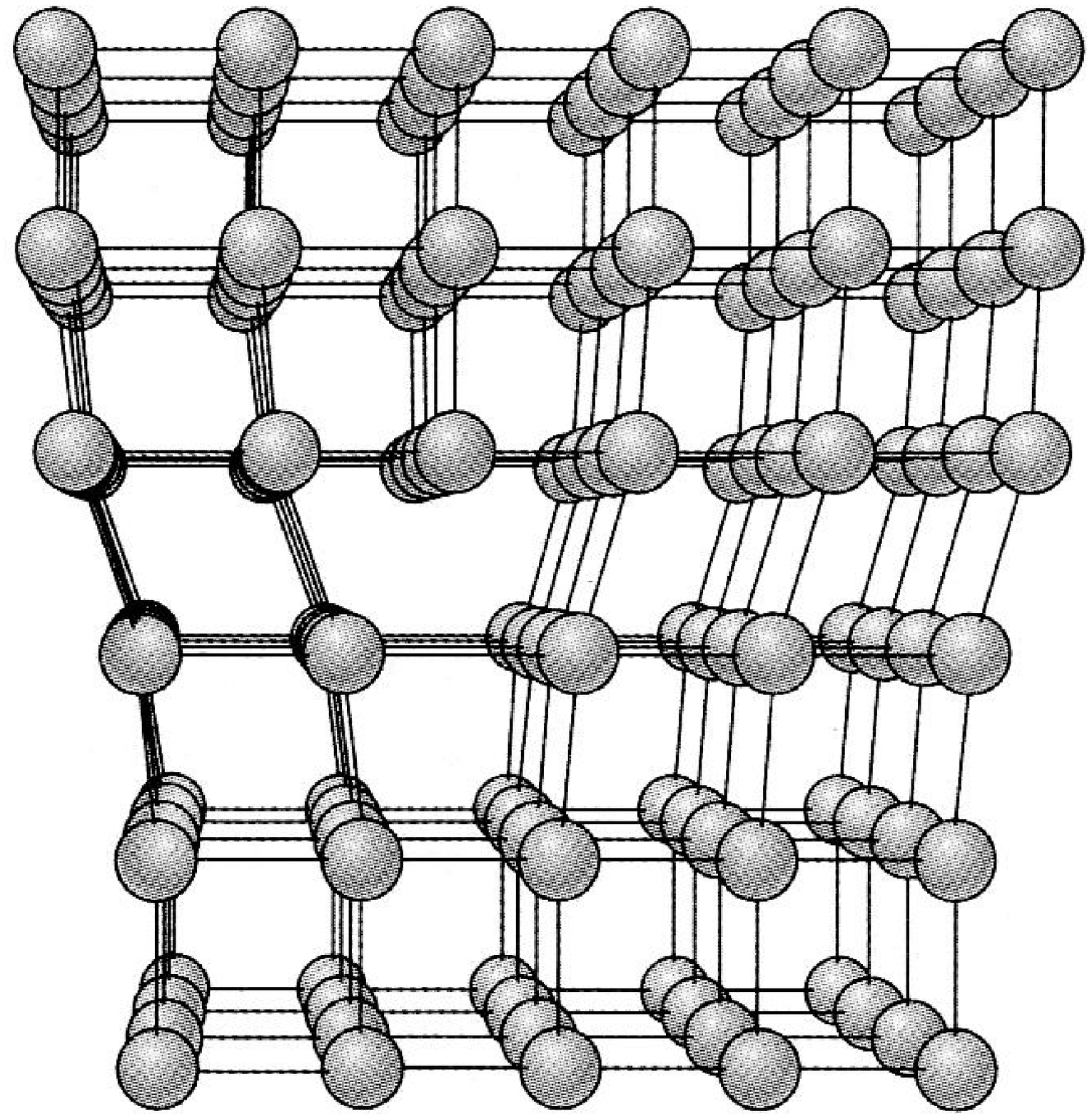}\hskip3em\includegraphics[width=5cm]{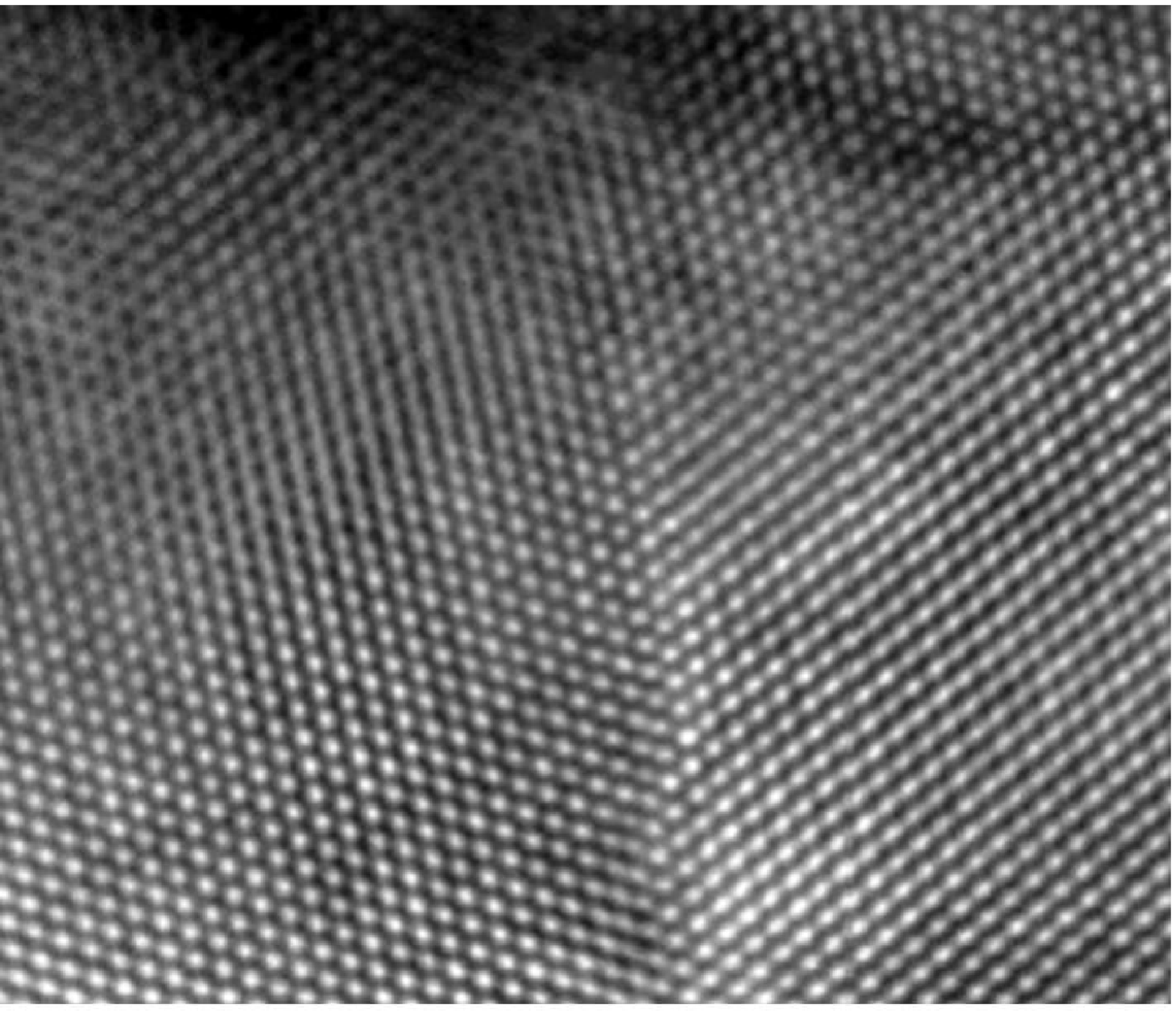}}\vskip.5em
\centerline{\includegraphics[width=13cm]{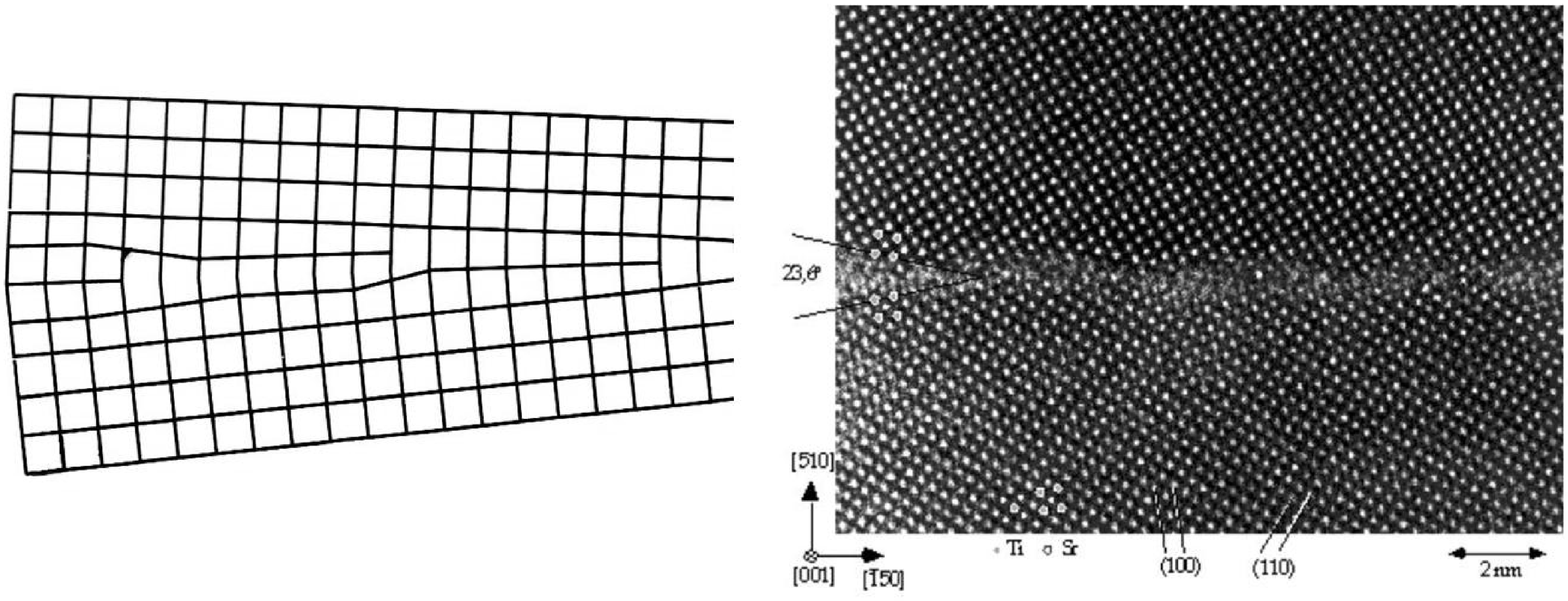}} \vskip.5ex
Figure 1: Edge dislocation and small angle grain boundary. The pictures on the left are
from [Ki] with kind permission of the publisher. The pictures on the right were taken with a TEM; references:
\begin{verbatim}http://www.fys.uio.no/bate/?page_\,id=7\end{verbatim}
\begin{verbatim}http://pruffle.mit.edu/~ccarter/NANOAM/images/\end{verbatim}\vskip1.5em
In this paper, we will use some results on a translational problem
to obtain spectral information about rotational problems in the
limit of small angles. Our main theorem deals with the following
situation. Let $V \colon \R^2 \to \R$ be a Lipschitz-continuous
function which is periodic w.r.t.\ the lattice $\Z^2$. For $\theta
\in (0,\pi/2)$, let
$$
M_\theta := \left(%
\begin{array}{cc}
  \cos\theta & -\sin\theta \\
  \sin\theta & \cos\theta \\
\end{array}%
\right)\in \R^{2 \times 2},
  \eqno{(0.2)}
$$
and
$$
V_\theta(x,y) := \left\{%
\begin{array}{lll}
  V(x,y), && \text{$x \ge 0$,} \\
  V (M_{-\theta}(x,y)), && \text{$x < 0$.} \\
\end{array}%
\right.\eqno{(0.3)}
$$
We then let $H_0$ denote the (unique) self-adjoint extension of
$-\Delta \restriction{\Cci{\R^2}}$, acting in the Hilbert space
 $L_2(\R^2)$, and
$$
R_\theta := H_0 + V_\theta, \qquad D(R_\theta) = D(H_0).
\eqno{(0.4)}
$$
Then $R_\theta$ is essentially self-adjoint on $\Cci{\R^2}$ and
semi-bounded from below. Our main assumption is that the periodic
Hamiltonian $H := H_0 + V = R_0$ has a gap $(a,b)$ in the
essential spectrum $\sigmaess(H)$, i.e., we assume that there
exist $a < b \in \R$ that satisfy $\inf\sigmaess(H) < a$ and
$(a,b) \cap \sigma(H) = \emptyset$; we do not need to assume that
$a$, $b$ are the actual gap edges. It is easy to see (using, e.g.,
[RS-I; Thm.~VIII.25]) that the operators $R_\theta$ converge
 to $R_{\theta_0}$ in the strong resolvent sense
as $\theta \to \theta_0 \in [0,\pi/2)$; in particular,
 $R_\theta$ converges to $H$ in the strong resolvent sense
as $\theta \to 0$. Recall that strong resolvent convergence
implies upper semi-continuity of the spectrum while the spectrum
may contract considerably when the limit is reached. In the
present paper, we are dealing with a situation where the spectrum
in fact behaves discontinuously at $\theta = 0$ since, counter to
first intuition, the spectrum of $R_\theta$ ``fills'' the gap
$(a,b)$ as $\theta \to 0$ with $\theta > 0$. This implies, in
particular, that  $R_\theta$ cannot converge to $H$ in the norm
resolvent sense, as $\theta \to 0$.
\vskip1em \noindent{\bf 0.1.~Theorem.} {\it Let $H$, $R_\theta$
and $(a,b)$ as above. Then, for any $\eps > 0$ there exists $0 <
\theta_\eps < \pi/2$ such that for any $E \in (a,b)$ we have }
$$
   \sigma(R_\theta) \cap (E-\eps, E+\eps) \ne \emptyset, \qquad \forall 0 <
   \theta < \theta_\eps.
   \eqno{(0.5)}
$$
\vskip.5em
\noindent{\bf 0.2.~Remarks.} {\parindent=1.5em
 \parskip=.5ex
$(i)$ Roughly speaking, the moment we start rotating the potential
on the left-hand side by a tiny angle the gap $(a,b)$ is suddenly
full of spectrum of $R_\theta$ in the sense that, for $0 < \theta
< \theta_\eps$, no gap of $R_\theta$ in the interval $(a,b)$ can
have length larger than $2\eps$. It is conceivable that for most
$\theta$ the spectrum of $R_\theta$ covers the interval $(a,b)$,
but there are examples (cf.\ Section 4) where $R_\theta$ has gaps
in $(a,b)$ for some $\theta$.

$(ii)$ It seems to be quite hard to determine the nature of the
spectrum of $R_\theta$ for general $\theta \in (0,\pi/2)$;
however, there are some  special angles for which a result from
[DS] excludes singular continuous spectrum (cf.\ Section 5).

$(iii)$ In addition to what is stated in Theorem 0.1 we obtain
lower and upper bounds for the spectral densities in the intervals
$(E-\eps, E+ \eps)$ on a scale that is appropriate to surface
states (without knowing that an integrated surface density of
states as in [EKSchrS, KS] exists for $R_\theta$); cf.\ Theorems
3.1 and 3.2.

} \vskip.5em There is a simple, intuitive connection between the
rotational problem and the related translational problem, given as
follows:
 Starting from the same periodic potential $V$ as above, we now look at
$$
    W_t (x,y) := \left\{%
\begin{array}{lll}
  V(x,y), && \text{$x \ge 0$,} \\
  V(x+t,y), && \text{$x < 0$,} \\
\end{array}%
\right.\qquad 0 \le 1 \le t, \eqno{(0.6)}
$$
 and define $D_t := -\Delta + W_t$, acting in $L_2(\R^2)$. In the 1-dimensional case,
 this problem has been studied in great detail by Korotyaev [Kor1, Kor2].
 A different approach was most recently implemented in
 [HK] where it is shown that some of the results
 of [Kor1, Kor2] can be recovered with a rather crude variational
 technique. This method can be easily generalized to the dislocation
 problem on a strip $\R \times (0,1)$ and then to the plane $\R^2$;
 one finds that spectrum of $D_t$ crosses the gap as $t$ varies between $0$ and $1$.
 Now our key observation consists in the following: for any given $\eps > 0$
and $n \in \N$, we can find points  $(0,\eta)$ on the $y$-axis
such that
$$
        |V_\theta(x,y) - W_t(x,y)| < \eps,
                      \qquad (x,y) \in Q_n(0,\eta),
  \eqno{(0.7)}
$$
with $Q_n(0,\eta) = (-n,n) \times (\eta-n, \eta+n)$,
 provided $\theta > 0$ is small enough and satisfies
a condition which ensures an appropriate alignment of the period
cells on the $y$-axis.  This basic observation is somewhat
reminiscent of a key idea in [HHK] where rotationally symmetric
Schr\"odinger operators of the type $-\Delta + U(|x|)$ in
$L_2(\R^n)$ with $U \colon \R \to \R$ periodic and
Lipschitz-continuous are considered: here, far away from the
origin, the potential $U(|x|)$ looks very much like a potential
depending only on the $x_1$-coordinate if we restrict our
attention to balls of fixed size with centers far out on the
$x_1$-axis.

The paper is organized as follows. In Section 1, we briefly
summarize some results of [HK] on translational lattice
dislocations for the strip  and for the plane. What we will use in
the sequel is the simple fact that,  for any $E \in (a,b)$, there
is some $t = t_E \in (0,1)$ with the following property: for any
$\eps > 0$,  there is a compactly supported approximate
eigenfunction $u$ in the domain of $D_t$ that satisfies
 $\norm{(D_t - E)u} < \eps$ and $\norm{u} = 1$.

In Section 2, we first employ the Birkhoff Ergodic Theorem to
obtain a set $\Theta\subset(0,\pi/2)$ with countable complement so
that (0.7) can be established for small $\theta \in \Theta$ and
suitable $\eta \in \R$. Then the above approximate eigenfunctions
$u$ will also be approximate eigenfunctions of $R_\theta$ for
small $\theta \in \Theta$, after an appropriate
($\theta$-dependent) translation along the $y$-axis. This then
gives  Theorem 0.1.

Suitable points $(0,\eta)$ for the construction of Section 2 occur
with a certain density and we expect a lower bound for the
integrated surface density of states. Since we do not know whether
the i.d.s.\ measure or the integrated surface density of states
measure exist, we only provide lower and upper bounds for the
number of Dirichlet eigenvalues in subintervals of the gap $(a,b)$
for our operators $R_\theta$, restricted to
 large squares $Q_n = (-n,n)^2$. Theorem 3.1 in Section 3 provides a lower
bound of the form $c_1 n$ for $n$ large with a positive constant
$c_1$, while Theorem~3.2 gives an upper bound by $c_2 n \log n$,
for $n$ large. Note that Theorem 3.2 deals with a much more
general situation: in fact, we allow here for two different
potentials $V_\ell$ and $V_r$ on the left and right which are not
required to be periodic; all we need is a common gap.

In Section 4 we discuss examples of {\it ``muffin tin''}-type
which come in three versions: in the simplest case, the muffin
tins are circular wells, arranged on a periodic grid, with
infinitely high walls (so that the Schr\"odinger operator is just
the direct sum of a countable number of Dirichlet Laplacians on
circles). We then approximate by muffin tins of finite height,
and, finally, by muffin tins with Lipschitz potentials. We obtain
spectral results for the rotation problem for all three versions.

In Section 5, finally, we first explain a simplified model for
small angle grain boundaries where we assume rotation by an angle
$\theta/2$ in both half-planes in opposite direction; this problem
is somewhat easier to deal with. We furthermore discuss special
rotation angles $\theta$ for which the operators $R_\theta$ are
periodic in the $y$-direction and thus have no singular continuous
spectrum [DS]. All these question hinge on an approximate or exact
matching up between the given periodic potential in the  right
half-plane and its rotated version in the left half-plane. This is
somewhat connected with the question of coincidence between a
lattice and some of its rotated versions (CSL-lattices).

For basic results on the spectral theory of self-adjoint operators
or, more specifically, periodic Schr\"odinger operators, we refer
to [K], [RS-IV], [E] and [Ku]. We will use results from these
sources mostly without specific reference.\\[.25cm]
{\it Acknowledgements.} The authors would like to thank Andreas
Ruschhaupt, Univ. Hannover, for several fruitful discussions and
the unknown referee for suggestions that helped to improve
the presentation.
\section{The dislocation problem on a strip
and for the plane}

Let $V \colon \R^2 \to \R$ be $\Z^2$-periodic and Lipschitz-continuous, let $I := (0,1)$, and let
$\Sigma := \R \times (0,1) = \R \times I$ denote the infinite strip of width
$1$.
 As above, we write $H:=-\Delta+V$ for the (self-adjoint) Schr\"odinger
operator with potential $V$ acting in $L_2(\R^2)$. Then
$\sigma(H)$, the spectrum of $H$, has band structure, i.e., it is the
(locally finite) union of compact intervals [RS-IV]. The intervals of
spectrum, the {\it bands}, may be separated by (open) intervals, the
{\it gaps}.  Moreover, $\sigma(H)$ is purely
absolutely continuous. For $0\leq t\leq 1$, we introduce the
self-adjoint operators
$$
\begin{array}{ccll}
  S_t  :=  -\Delta+W_t, &\hbox{acting in } L_2(\Sigma), \\
  D_t  :=  -\Delta+W_t, &\hbox{acting in } L_2(\R^2),   \\
\end{array}\eqno{(1.1)}
$$
 where $S_t$ has periodic boundary conditions in the $y$-variable and $W_t$ is as in (0.6).
 Since $V$ is bounded, the domains $D(.)$ of the above operators satisfy
 $D(D_t) = D(H)$ and $D(S_t) = D(H_{0,\Sigma})$, for all $t$,
where $H_{0,\Sigma}$ denotes the Laplacian on $\Sigma$ with periodic boundary conditions in
$y$.  The operator $-\Delta+W_t$ in $L_2(\Sigma)$ with
 $\theta$-periodic boundary conditions in $y$ is denoted by $S_t(\theta)$,
for $0 \le \theta \le 2\pi$.
 As usual, $D_t$ can be obtained from the $S_t(\theta)$ as a direct
 fiber integral,
 $$
      D_t = \int_{0\le \theta \le 2\pi }^\oplus S_t(\theta) \frac{d\theta}{2\pi};
    \eqno{(1.2)}
 $$
 direct fiber integrals are discussed, e.g., in [RS-IV, DS].
 As a consequence, for any $\theta$ the spectrum of
 $S_t(\theta)$ is a subset of $\sigma(D_t)$. Furthermore, using the
 periodicity in the $x$-direction, each $S_t(\theta)$ can itself be written as
 a direct fiber integral and so the spectrum of $S_t(\theta)$
 is purely essential spectrum with a band-gap structure.

 We now give a condensed account of some of the results in [HK] concerning
 the operators $S_t$ and $D_t$. We begin with $S_t$ where we first note
 the following well-known basic facts:

(1) Adding in a Dirichlet boundary condition on any given vertical
 line segment $\{(x,y) \in \Sigma \: ; \: x = \xi \}$, for some $\xi \in \R$,
 leads to a compact perturbation of the resolvent of $S_t$,
 for all $t$. In fact, if  $H_{0,\Sigma,D}$ is defined as
 $H_{0,\Sigma}$ above, but now with an additional Dirichlet boundary condition
 at $x = \xi$, one can write down explicit formulae for the integral kernels
 of the resolvents by using the reflection principle. One finds that the
 operator $(H_{0,\Sigma} + 1)^{-1} - (H_{0,\Sigma,D} + 1)^{-1}$ is Hilbert-Schmidt.
 It is then easy to show that $(H_{0,\Sigma} + W_t + c)^{-1} - (H_{0,\Sigma,D}
 + W_t + c)^{-1}$ is compact for all sufficiently large $c \ge 0$.

(2) It follows from (1) that $\sigmaess(S_t) = \sigmaess(S_0)$. To see this,
just add in Dirichlet boundary conditions at $x = 0$ into $S_0$ and $S_t$,
and another Dirichlet boundary condition at $x = -t$ into $S_t$ to obtain
the operators $S_{0,D}$ and $S_{t,D}$. Then the
parts of $S_{0,D}$ and $S_{t,D}$ on $(0,\infty) \times I$ are equal while the
contribution to $S_{t,D}$ from $(-\infty, -t) \times I$ is unitarily equivalent to the
part of $S_{0,D}$ in $(-\infty,0) \times I$. Finally, the part of $S_{t,D}$ associated
with $(-t,0) \times I$ has compact resolvent.

(3) The eigenvalues of $S_t$ in the gaps of $S_0$ are continuous functions of
the parameter $t$. (A simple proof can be obtained by scaling the interval $(-n-t,n)$ in such a
way that we can use $L_2((-n,n) \times I)$ as a common Hilbert space).

\vskip1em
The following result shows that there is non-trivial spectral flow
of the family $S_t$ through the gaps of $S_0$:
\vskip1em
{\bf 1.1.~Proposition.} {\it Let $(a,b)$ denote a
spectral gap of $H$ and let $E \in (a,b)$. Then there exists some $t =
t_E \in (0,1)$ such that $E$ is a (discrete) eigenvalue of $S_{t_E}$.

 Moreover, for any $n \in \N$ there are functions $v_n = v_n(x,y)$ in the domain of $S_t$
 that satisfy $\norm{v_n}=1$, $\supp v_n\subset[-n,n] \times [0,1]$ and
 $(S_{t_E} - E)v_n\to 0$ as $n\to\infty$.
}
\vskip1em

For the proof of Proposition 1.1, we use an approximation by operators with compact resolvent
on finite sections of $\Sigma$ (the basic idea is somewhat reminiscent of [DH, ADH]):
  For $n \in \N$, let $S^{(n)}_t$ denote the operators $-\Delta + W_t$ on
$\Sigma^{(n)}_t := (-n-t,n) \times I$ with periodic boundary conditions in both variables.
 The spectrum of the operators $S^{(n)}_t$
is purely discrete.  Again, the eigenvalues of $S^{(n)}_t$ depend continuously on $t$.
 Furthermore, routine arguments from Floquet-theory
 imply that

$(i)$ $\sigma(S^{(n)}_0) \subset \sigma(S_0)$, for all $n \in \N$,

\noindent and

$(ii)$ for any gap $(a,b)$ of $S_0$ there exists a fixed number $m \in \N$
such that the operators $S^{(n)}_0$ (respectively, $S^{(n)}_1$) have precisely
 $2 n m$ (respectively, $(2n+1)m$) eigenvalues below $a$, counting
 multiplicities (cf., e.g., [RS-IV; proof of Thm.~XIII.101] or [E;
 Thm.~6.2.1]).
\vskip1ex

As $t$ increases from $0$ to $1$, the operators $S^{(n)}_0$ are transformed into
$S_1^{(n)}$. It is now immediate from property $(ii)$ and the continuity of the
eigenvalues that, as $t$ grows from $0$ to $1$, at least one
eigenvalue of $S^{(n)}_t$ must cross $E$.


We have thus shown that,  for each $n \in \N$,
there is some $t_n \in (0,1)$ and an eigenfunction $u_n \in D(S_{t_n}^{(n)})$ such
that $\norm{u_n} = 1$ and $S_{t_n}^{(n)} u_n = E u_n$. Since the $u_n$ obey
periodic boundary conditions with respect to the $x$-variable on
$\Sigma_{t_n}^{(n)} = (-n-t_n,n) \times I$,
  we can use routine arguments to show that the parts of $u_n$ (and of $\nabla u_n$)
 outside of $(-n/2,n/2)$ go to zero, as $n \to \infty$. Indeed, let us fix
cut-off functions  $\phi_n = \phi_n(x) \in C_c^\infty(-n/2,n/2)$ that
satisfy $\phi_n(x) = \phi_1(x/n)$ and $\phi_1(x) = 1$ for $-1/4 \le x \le 1/4$.
Then $r_n := (1 - \phi_n) u_n$ vanishes in $(-n/4,n/4)\times I$.
 We now define ${\tilde r}_n$ by ${\tilde r}_n(x,y) := r_n(x,y)$, for $0 < x < n$,
and ${\tilde r}_n(x,y) := r_n(x - 2n - t_n,y)$, for $x > n$, i.e., we
translate the part of $r_n$ in $(-n-t_n,0) \times I$ to the right by
$2n + t_n$. Since $r_n$ satisfies periodic boundary conditions, it is
clear that ${\tilde r}_n$ belongs to the domain of the periodic operator $S_0$
 and that $(S_0 - E) {\tilde r}_n \to 0$, as $n \to \infty$. As $E$ is
in a gap of $S_0$, this implies ${\tilde r}_n \to 0$ and then also $r_n \to
0$.
We therefore see that $v_n := \phi_n u_n$ satisfies
$$
   \norm{v_n} \to 1, \qquad (S^{(n)}_{t_n} - E) v_n \to 0,
   \eqno{(1.3)}
$$
as $n \to \infty$. Without loss of generality we may assume that the $t_n$
converge to some $t = t_E \in (0,1)$. But $v_n \in D(S_{t_n})$ and $S_{t_n} v_n = S^{(n)}_{t_n}v_n$,
and we find that $(S_{t_E} - E) v_n \to 0$, as $n \to \infty$ (recall that
$V$ is Lipschitz-continuous). Now the Spectral Theorem implies $E \in \sigma(S_{t_E})$,
as required.
Furthermore, the functions $v_n$ clearly enjoy the property stated in the second part
of Proposition 1.1.

The functions $v_n$ constructed above satisfy periodic
boundary conditions with respect
to $y$ and may thus be extended to $y$-periodic functions ${\tilde v}_n$ on
$\R^2$. Applying also cut-offs $\psi_n = \phi_n(y)$ in the $y$-direction, we let
$$
    w_n := \frac{1}{\norm{\psi_n {\tilde v}_n}} \psi_n {\tilde v}_n;
   \eqno{(1.4)}
$$
the $w_n$ satisfy $\norm{w_n} =1 $, $w_n \in D_{t_n}$ and $(D_{t_n} - E)w_n
\to 0$ as $n \to \infty$. By the same argument as above this leads to
$E \in \sigma(D_{t_E})$ (where, again $t_E = \lim t_n$) and
we have thus obtained:

\vskip1.5em
{\bf 1.2.~Proposition.} {\it Let $(a,b)$ denote a
spectral gap of $H$ and let $E \in (a,b)$. Then there exists $t =
t_E \in (0,1)$ such that $E \in \sigma(D_{t_E})$.

 Moreover, for any $n \in \N$ there are functions $w_n=w_n(x,y)$ in the domain
 of $D_t$ that satisfy $\norm{w_n}=1$, $\supp w_n\subset[-n,n]^2$ and
 $(D_{t_E}-E)w_n\to 0$ as $n\to\infty$.
}
\vskip1em
Note that the spectrum of $D_t$ inside $(a,b)$ will again consist of bands
 which we could find by repeating the above process for all $\theta$-periodic
boundary conditions w.r.t.\ $y$.
For a detailed discussion and further results, we refer to [HK].
%
%
\section{The rotation problem for small angles}
 In this section, we study the spectrum of the operators
 $R_\theta$, for $0 < \theta < \pi/2$, where the $R_\theta$ are defined in (0.4)
  as self-adjoint operators in the Hilbert space $L_2(\R^2)$.

 In view of a proof of Theorem 0.1, consider a fixed $E \in (a,b)$. Then, by
 Proposition 1.1, there  is some  $t \in (0,1)$ such that $E$ is in the spectrum of the dislocation
 operator $D_t$ on the plane.
 We wish to find angles $\theta$ with the property that the
 potential $V_{\theta}$ is approximately equal to $W_t$ on a
 sufficiently large square $Q_n(0,\eta)$ of side-length $2n$, centered at some point
 $(0,\eta)$ on the $y$-axis. This leads to the following requirements:
 If we imagine the grid $\Gamma = \{(x,y) \in \R^2 \: ; \: x \in \Z \hbox{\rm\ or\
 } y \in \Z \}$  of lines describing  the period cells,  we have to make sure
 that, inside $Q_n(0,\eta)$, the  alignment between the horizontal lines of
 $\Gamma$ in the right half-plane with the rotated horizontal lines of
 $M_\theta\Gamma$ in the left half-plane is nearly perfect on the $y$-axis
 and that the rotated vertical lines of $M_\theta \Gamma$ in the left half-plane
 have, roughly, distance  $t$ (modulo $\Z$) from the $y$-axis.
 More precisely, we wish to find $m \in \N$ such
 that $m/\cos\theta$ is integer, up to a small error, and $m \tan\theta = t\,(\text{mod }\Z)$,
 again up to a small error, inside $Q_n(0,\eta)$.

We first prepare a lemma which deals with ergodicity on the flat
torus $\T^2 = \R^2 / \Z^2$, as in [RS-I], [CFS]. We consider
transformations $T_\theta \colon \T^2 \to \T^2$ defined by
$$
      T_\theta(x,y):=(x+\tan\theta,y+1/\cos\theta).
    \eqno{(2.1)}
$$
\vskip.5em\noindent {\bf 2.1.~Lemma.} {\it There is a set $\Theta
\subset (0,\pi/2)$ with countable complement such that the
transformation $T_\theta$ in $(2.1)$ is ergodic for all $\theta
\in \Theta$.} \vskip1em
\noindent{\bf Proof.} $T_{\theta}$ is ergodic if and only if the
numbers $1$, $\tan\theta$, and $1/\cos\theta$ are independent over
the rationals, i.e., $(n_1,n_2,n_3) \in \Z^3$ and
$$
        n_1 + n_2\tan\theta + \frac{n_3}{\cos\theta} = 0
 \eqno{(2.2)}
$$
implies $n_1 = n_2 = n_3 = 0$. Write ${\mathcal Z}_3 := \Z^3
\setminus \{(0,0,0)\}$. For any triple $(n_1, n_2, n_3) \in
{\mathcal Z}_3$ the set of points $(x,y) \in \R^2$ that satisfy
$n_1 + n_2 x + n_3 y = 0$ is a line
 $\ell = \ell_{(n_1,n_2,n_3)} \subset \R^2$. Consider the (countable) set
$$
   \Lambda := \{ \ell_{(n_1,n_2,n_3)} ; (n_1, n_2, n_3) \in {\mathcal Z}_3 \} .
    \eqno{(2.3)}
$$
In (2.2), the variables $x$ and $y$ are of the special form $x =
\tan\theta$, $y = 1/\cos\theta$ and so  $y = \sqrt{1 + x^2}$.
Since  $F(x):= \sqrt{1+x^2}$ is convex, each $\ell \in \Lambda$ has
at most two intersection points with the graph $G(F)$ of $F$.
  Then
$$
         S:=\cup_{\ell\in\Lambda}\{\ell\cap G(F)\}
  \eqno{(2.4)}
$$
is countable and so $G(F) \setminus S$ has full $1$-dimensional measure.
 Let $G(F)_+ := G(F) \cap \{ x > 0 \}$.
 We map $G(F)_+\backslash S$ to $(0,\pi/2)$ by
$h \colon (x,F(x))\mapsto\arctan x$. Since $h \colon G(F)_+ \to
(0,\pi/2)$ is diffeomorphic, $\Theta:=h(G(F)_+\backslash S)$ is
as desired.\hfill$\square$ \vskip1.5em
Let us write $x_\sim$ for the fractional part of $x > 0$, i.e.,
$x_\sim=x-\lfloor x\rfloor$ if $x>0$. In the proof of our main
theorem, we will need natural numbers $m$ such that, for
$t\in(0,1)$  given, $(m\tan\theta)_\sim$ is approximately equal to
$t$ and $(m/\cos\theta)_\sim$ almost equals $0$. The existence of
such numbers $m$ follows from Lemma 2.1 and Birkhoff's Ergodic
Theorem. Let $\theta\in\Theta$, $\eps>0$, and let us denote by
$\chi_Q$ the characteristic function of the set
$Q:=(t-\eps,t+\eps)\times(-\eps,\eps)\subset\T^2$. Then, for all
$(x,y)\in\T^2$,
$$
       \lim_{n\to\infty}\frac{1}{n}\sum_{m=0}^{n-1}\chi_Q
       (T_\theta^m(x,y))=\int_{Q}\dx \dy =4\eps^2>0,
  \eqno{(2.5)}
$$
and we may take $(x,y) := (0,0)$ to arrive at the desired result.

We add the following remarks to the above argument:

(1) Translation on the torus is a particularly simple ergodic transformation:
for $\theta$ given, it can equivalently be seen as
linear motion on parallel lines in $\R^2$,  factored by $\Z^2$. In particular,
two nearby points $(x,y)$ and $(x',y')$ will forever keep their relative
position under the action of $T_\theta^m$, and
thus the statement of Birkhoff's Theorem holds for {\it any} point $(x,y)$, not
just for a.e.\ $(x,y)$ (cf., e.g., [CFS; Ch.~3, Par.~1]).

(2) In some sense, the Birkhoff Theorem is the strongest result one can use
in this context. Similar results are obtained from Dirichlet's Theorem
on the approximation of irrational numbers by rationals.

\vskip1em

We are now ready for a first main result which establishes the
existence of surface states in the gaps of $H$ and shows that, in
fact, any gap $(a,b)$ of $H$ is filling up with spectrum of
$R_\theta$ as $\theta \to 0$. \vskip1.5em

\noindent{\bf 2.2.~Proposition.} {\it Let $(a,b)$ be a spectral
gap of $H$ and let $[\alpha, \beta] \subset (a,b)$, $\alpha <
\beta$. Then there is a $\theta_0=\theta_0(\alpha,\beta) > 0$ such
that}
$$
   \sigma(R_\theta) \cap (\alpha,\beta) \ne \emptyset, \qquad
   \forall\theta\in(0,\theta_0).
   \eqno{(2.6)}
$$
 \vskip1em
\noindent{\bf Proof.}
(1) We first restrict our attention to $\theta \in \Theta$ with
$\Theta$ as in Lemma 2.1. Let $E \in (\alpha,\beta)$ and $\eps :=
\min\{E-\alpha, \beta - E\}/2$. By Proposition 1.1, we can find
$n=n_{\eps}\in\N$ and a function $u_n$ of norm 1 in the domain of
$D_t$
 with $\supp u_n\subset[-n,n]^2$ such that $\norm{(D_t-E)u_n}<\eps$. Obviously
$u_{n,k}(x,y):=u_n(x,y-k)$ satisfies the same estimate for any
$k\in\N$. If we can show that, for appropriate $k\in\N$,
$$
     |V_\theta(x,y) - W_t(x,y)| < \eps, \qquad (x,y) \in Q_n(0,k)
 \eqno(2.7)
$$
(recall the definition of $Q_n(0,k) = (-n,n) \times (k-n,k+n)$),
we may conclude that
$$
        \norm{(R_\theta-E)u_{n,k}}< 2\eps;
  \eqno{(2.8)}
$$
but then the Spectral Theorem implies that $R_\theta$ has spectrum
inside the interval $(E - 2 \eps, E + 2\eps) \subset
(\alpha,\beta)$.

For a proof of (2.7), we first observe that by the properties of
$V$ and the definitions of $V_\theta$ and $W_t$, we have the
following estimate:
$$
  \left|V_\theta(x,y) - W_t(x,y) \right|^2\leq
   \min_{j_1,j_2 \in \Z} L^2 ((X-j_1)^2 + (Y -
   j_2)^2),\quad\forall(x,y)\in\R^2,
\eqno{(2.9)}
 $$
with
$$
  X := x (\cos \theta - 1) - t + y \sin\theta,
\qquad Y := -x  \sin \theta + y (\cos\theta - 1)
  \eqno{(2.10)}
$$
and $L$ the Lipschitz constant of $V$. Now for $\theta\in \Theta$
given, there is some $m = m_\theta \in \N$ such that
$$
  \left( \frac{m}{\cos\theta}\right)_\sim < \eps/4, \qquad
  |(m \tan \theta)_\sim - t| < \eps/4;
  \eqno{(2.11)}
$$
in particular, there is some $N \in \N$ s.th.\ $|m/\cos\theta - N|
< \eps/4$.

We may now apply the estimate $(2.9)$ to the points $(x,y) \in
Q_n(0,N)$ to find
$$
  \left| V_\theta(x,y) - W_t(x,y)\right|^2 \le L^2
  \left( (X - \lfloor m \tan\theta \rfloor)^2 + (Y + N - m)^2 \right),
  \eqno{(2.12)}
$$
for all $(x,y) \in Q_n(0,N)$. Here
$$
  | X - \lfloor m \tan\theta \rfloor | \le n (1 - \cos\theta) +
    n \theta + | m \tan\theta - \lfloor m \tan \theta \rfloor - t|
   \le 2 n_\eps \theta + | (m \tan \theta)_\sim - t|
    \eqno{(2.13)}
$$
and
$$
     |Y+N-m|\leq 2n_\eps\theta+|N-m/\cos\theta|.
  \eqno{(2.14)}
$$
We choose $\theta_0 > 0$ small enough to have $2 n_\eps \theta_0 <
\eps/4$ and $(2.7)$ follows if we pick $k := N$. We have thus
shown that $R_\theta$ has spectrum in $(\alpha,\beta)$ for all
$\theta \in \Theta \cap (0,\theta_0)$.

(2) In order to remove the restriction $\theta \in \Theta$ we note
that
 with each $\theta\in\Theta$ there comes a  positive number $\eta_\theta>0$ such that
$$\norm{(R_\sigma-E)u_{n,k}}<3\eps,\quad\forall\sigma\in(\theta-\eta_\theta,\theta+\eta_\theta),
  \eqno{(2.15)}
$$
since
$$
      \norm{(V_\sigma - V_\theta) \restriction \supp u_{n,k}}_\infty \to
      0,\quad\sigma\to\theta.
    \eqno{(2.16)}
$$
As the intervals $(\theta-\eta_\theta,\theta+\eta_\theta)$ with
$\theta$ ranging between $0$ and $\theta_0$ cover the interval
$(0,\theta_0)$, the desired result follows.
\hfill$\square$\vskip1.5em
Now it is easy to obtain Theorem 0.1 in the Introduction from
Proposition 2.2: \vskip1em

\noindent{\bf Proof of Theorem 0.1.} For $\eps>0$ given, we
consider points $a=\gamma_0<\gamma_1<\gamma_2<\ldots<\gamma_N=b$
such that $\gamma_j-\gamma_{j-1}<\eps/2$, for $j=1,\ldots,N$. For
each of the intervals $I_j:=(\gamma_{j-1},\gamma_{j})$, $2\leq
j\leq N-1$, Proposition 2.2 yields a constant $\theta_j>0$ with
the property that $R_\theta$ has spectrum in the interval $I_j$
for all $0<\theta<\theta_j$. Then $\theta_0:=\min_{2\leq j\leq
N-1}\theta_j$ has the required properties. \hfill$\square$
\vskip1.5em
\section{Integrated density of states bounds}
It is clear that ergodicity gives us not just a single $m$ as in
 (2.11), for $\theta \in \Theta$; in fact, eqn.~(2.5) guarantees
 that suitable $m$ will appear with a certain frequency. We will use
 this observation to obtain lower bounds for a quantity which,
 in the limit, would translate into a (positive) lower bound for
 the surface i.d.s.\ measure if we knew that the required limit
 exists. This will be complemented by a similar upper bound
 which is of the expected order, up to a logarithmic factor.
 A detailed and rather complete account of the i.d.s.\ for (random)
 Schr\"odinger operators can be found in [V] which also contains
 a wealth of references. [EKSchrS] and [KS] specifically  discuss
 the existence of a surface i.d.s.\ as a distribution or a
 measure. Some results on the surface i.d.s.\ measure
 for the translational dislocation problem can be found in [HK].

  Let $R^{(n)}_\theta$ denote the operator $-\Delta +  V_\theta$,
  acting in $L_2(Q_n)$ with Dirichlet boundary
 conditions, where $Q_n := (-n,n)^2 \subset \R^2$. For any
 interval $I \subset \R$, we denote by $N_I(R^{(n)}_\theta)$ the
 number of eigenvalues of $R^{(n)}_\theta$ in  $I$,
 each eigenvalue being counted according to its multiplicity.
 The existence of a surface i.d.s.\ measure in the gap $(a,b)$
 would correspond to the existence
 of a finite limit $\lim_{n\to\infty} \frac{1}{n} N_I(R^{(n)}_\theta)$,
 for any interval $I$ with $\overline{I} \subset (a,b)$.
 Theorem 3.1 provides lower bounds of the form
 $$
      \liminf_{n\to\infty} \frac{1}{n} N_I(R^{(n)}_\theta) > 0,
    \eqno{(3.1)}
 $$
 for (non-degenerate) subintervals $I$ and small $\theta \in \Theta$,
 while Theorem 3.2 yields an upper bound
 $$
      \limsup_{n\to\infty} \frac{1}{ n \log n} N_I(R^{(n)}_\theta) < \infty.
   \eqno{(3.2)}
 $$
We begin with a lower bound.
\vskip1.5em\noindent {\bf 3.1.~Theorem.}
 {\it Let $H$, $R_\theta$ as above and suppose that
$(a,b)$ is a spectral gap of $H$. Let $\Theta$ as in Lemma 2.1.

Then, for any $\eps > 0$ there exists a $\theta_\eps > 0$ such
that $(3.1)$ holds for all $\theta \in \Theta \cap
(0,\theta_\eps)$ and for any interval $I \subset (a,b)$ of length
greater than $\eps$.}

\vskip1em

\noindent{\bf Proof.} (1) Let $[\alpha,\beta] \subset (a,b)$, fix
$E \in (\alpha,\beta)$, and let $0 < \eps <  \min\{ E - \alpha,
\beta - E \}$. Let $u_0$ in the domain of $D_t$ with compact
support satisfy $\lnorm{u_0} = 1$  and $\norm{(D_t - E) u_0} <
\eps$,
 as in  Prop.\ 1.1. Let $\nu \in \N$ be such that
$\supp u_0 \subset Q_\nu = (-\nu,\nu)^2$; note that, in this
proof, $\nu$ corresponds to the parameter $n$ of Section 2.

 Let $\theta \in \Theta \cap (0,\pi/4]$ so that, in particular, $1/\sqrt{2}
 \le \cos\theta \le 1$. By ergodicity, there exists a constant
 $c_0 = c_0(\theta)> 0$ with the following properties: for $n \in \N$ large, there are
 at least $J_n : = \lfloor c_0 n \rfloor$ natural numbers $m_1, \ldots,
 m_{J_n} \in (0, n/4)$ such that (2.11) holds for $m = m_s$, $s=1, \ldots, J_n$,
 and such  that
$$
 |m_s - m_r| \ge 2\nu, \qquad s \ne r, \quad 1 \le s, r \le  J_n;
  \eqno{(3.3)}
$$
 here $J_n$ and $m_1, \ldots, m_{J_n}$ depend on $n$ and $\theta$.
 It follows that for each $j = 1, \ldots, J_n$ there is some $N_j \in \N$
 such that $|m_s/\cos\theta - N_j| < \eps/4$ and
 $|m_s \tan\theta - t|_\sim < \eps/4$. We then see that the functions
 $\phi_j$, defined by  $\phi_j(x,y) := u_0(x, y - N_j)$,
 are of norm 1 and have  mutually disjoint supports contained in
 $(-n,n)^2$. Furthermore, for $\theta$ small enough, $0 < \theta <
 \theta_\eps$, say, we can show (as in the
 proof of Proposition 2.2) that an estimate (2.7) holds on
 each square $(-\nu,\nu) \times (N_j - \nu, N_j + \nu)$. Thus (2.7) holds on
 the support of each $\phi_j$ and it follows that
 $$
     \norm{(R_{\theta}^{(n)}-E)\phi_j}<\eps, \qquad
      0 < \theta < \theta_\eps, \quad j = 1, \ldots, J_n.
    \eqno{(3.4)}
$$
Then $\MM:=\hbox{span}\set{\varphi_j}{j=1,\ldots,J_n}$ has
dimension $J_n$. Let $\NN$ denote the range of the spectral
projection $P_{(\alpha,\beta)}(R_\theta^{(n)})$ of
$R_\theta^{(n)}$ associated with the interval $(\alpha,\beta)$ and
assume for a contradiction that $\dim \NN < J_n$. Then we can find
a function $v \in \MM \cap \NN^\perp$ of norm 1. By the Spectral
Theorem,
 $\norm{(R_\theta^{(n)} - E)v} \ge \eps$. On the other hand, (3.4) and
 $v=\sum_{i=1}^Na_i\varphi_i$ implies $\norm{(R_\theta^{(n)} - E)v} < \eps$
 because the $\phi_j$ have mutually disjoint supports.

We  have therefore shown that for any interval $I= [\alpha,\beta]$
there exists some $\theta_0 > 0$ such that (3.1) holds for all
$\theta \in \Theta \cap (0,\theta_0)$. \vskip.5ex (2) Now let
$\eps > 0$. As in the proof of Theorem 0.1, given at the end of
Section 2, we may cover the interval $(a,b)$ by a finite number of
subintervals of length $\eps$; applying the result of part (1) we
then obtain the desired statement.
\endproof
\vskip1em \noindent
{\bf Remarks.} (a) It appears that the
argument used at the end of the proof of
 Proposition 2.2 to  remove the restriction $\theta \in \Theta$ does not
 work in the context of Theorem 3.1.

 (b) It follows from the proof of Thm.~3.1 that $\sigmaess(R_\theta) \cap I
 \ne \emptyset$ for all  $\theta \in \Theta \cap (0,\theta_\eps)$ and
 for any interval $I \subset (a,b)$ of length greater than $\eps$.

\vskip1.5em

We now complement the lower estimate established in Theorem 3.1 by
an upper bound which is of the expected order, up to a logarithmic
factor. Note that we treat a situation which is far more general
than the rotation or dislocation problems studied so far. In fact,
we will allow for different potentials $V_1$ on the left and $V_2$
on the right which are only linked by the assumption that there is
a common spectral gap; neither $V_1$ nor $V_2$ are required to be
periodic. The proof uses technology which is fairly standard and
based on exponential decay estimates for resolvents.
\vskip1.5em
\noindent {\bf 3.2.~Theorem.} {\it Let $V_1$, $V_2 \in
L_\infty(\R^2,\R)$ and suppose that the interval $(a,b) \subset
\R$ does not intersect the spectra of the self-adjoint operators
$H_k := -\Delta + V_k$, $k = 1, 2$, both acting in the Hilbert
space $L_2(\R^2)$. Let
$$
  W := \chi_{\{x<0\}} \cdot V_1 + \chi_{\{x \ge 0\}} \cdot V_2
  \eqno{(3.5)}
$$
and define $H:=-\Delta+W$, a self-adjoint operator in $L_2(\R^2)$.
Finally, we let $H^{(n)}$ denote the self-adjoint operator
$-\Delta+W$ acting in $L_2(Q_n)$ with Dirichlet boundary
conditions.
Then, for any interval $[a',b'] \subset (a,b)$, we have
$$
  \limsup_{n\to\infty}\frac{1}{n\log n}N_{[a',b']}(H^{(n)}) < \infty.
 \eqno{(3.6)}
$$
}
\vskip.5em
\noindent {\bf Proof.}
 (1) We write $N(n) := N_{[a',b']}(H^{(n)})$ and note that there is
a constant $c_0 \ge 0$ such that
$$
        N(n) \le c_0 n^2, \qquad n \in \N;
  \eqno{(3.7)}
$$
this follows by routine min-max arguments as in [RS-IV; Section
XIII.15].

(2) Let us consider the (normalized) eigenfunctions $u_{i,n}$ of
$H^{(n)}$ associated with the eigenvalues $E_{i,n} \in [a',b']$,
for $i = 1, \ldots, N(n)$. The main idea of the proof is to show
that the $u_{i,n}$ are concentrated near the boundary of $Q_n$ or
near the $y$-axis. To obtain the corresponding estimates, we
introduce
 the sets
$$
\Omega_j(n) :=\Omega_j^-(n)\cup\Omega_j^+(n), \qquad j \in \{1, 2,
3, 4 \},
 \eqno{(3.8)}
$$
where $\Omega_j^-(n) := \left(-\frac{n}{2}+\frac{2j}{\alpha}\log
n,-\frac{2j}{\alpha}\log n\right)\times
\left(-\frac{n}{2}+\frac{2j}{\alpha}\log
n,\frac{n}{2}-\frac{2j}{\alpha}\log n\right)$, and $\Omega_j^+(n)
:= - \Omega_j^-(n)$ is the mirror-image of $\Omega_j^-(n)$ with
respect to the $y$-axis; the parameter $\alpha > 0$ will be chosen
as in eqn.~(3.9) below. Note that, for $\alpha > 0$ fixed, the
sets $\Omega_1{(n)}, \ldots, \Omega_4(n)$ are non-empty for $n$
large. We have the trivial inclusions $\Omega_{j+1}(n) \subset
\Omega_j(n)$ for $j = 1, 2, 3$.

 We will use the following exponential decay
 estimate for the resolvent of the operators $H_k$: There are constants  $C \ge 0$,
 $\alpha>0$ such that for any $E\in [a',b']$ and (measurable)
 sets $K_1,K_2\subset\R^2$ we have (cf., e.g., [AADH; Prop.~2.4])
$$
  \norm{\chi_{K_1}\partial_j^p(H_k - E)^{-1}\chi_{K_2}}
  \leq C \hbox{\rm e}^{-\alpha\,\text{dist}(K_1,K_2)}, \quad j, p \in
  \{0,1\}, \quad k \in \{1, 2\};
  \eqno{(3.9)}
$$
here $\partial_1 = \partial_x$, $\partial_2 = \partial_y$.
 We also choose cut-off functions $\varphi_n,\psi_n\in
 C_c^{\infty}(\R^2;\R)$ satisfying
$$
  \text{supp}\,\varphi_n\subset\Omega_1(n),\quad
  \varphi_n\upharpoonright{\Omega_2(n)}=1,\quad
  \text{supp}\,\psi_n\subset\Omega_3(n), \quad
  \psi_n\upharpoonright{\Omega_4(n)}=1,
 \eqno{(3.10)}
$$
and
$|\nabla\phi_n|,|\nabla\psi_n|,|\partial_{ij}\phi_n|,|\partial_{ij}\psi_n|\leq
c(\log n)^{-1}$ with some constant $c \ge 0$; here $\phi_n =
\phi_{n, \ell} + \phi_{n,r}$ with $\phi_{n,\ell}$ and $\phi_{n,r}$
being supported in $\Omega_1^-(n)$ and $\Omega_1^+(n)$,
respectively. By a well-known argument we can now derive the
desired localization property: by the Leibniz rule, we have for $i
= 1, \ldots, N(n)$
$$
   (H_1 - E_i)(\phi_{n,\ell}u_{i,n})=(H^{(n)}-E_i)(\phi_{n,\ell}u_{i,n})=
        -2\nabla\phi_{n,\ell}\cdot\nabla u_{i,n}-\Delta\phi_{n,\ell}u_{i,n}
 \eqno{(3.11)}
$$
so that
$$
  \chi_{\Omega_3^-(n)}u_{i,n}=-\chi_{\Omega_3^-(n)}(H_1 - E_i)^{-1}\chi_{\text{supp}\nabla\phi_{n,\ell}}
[2\nabla\phi_{n,\ell}\cdot\nabla
u_{i,n}+\Delta\phi_{n,\ell}u_{i,n}].
 \eqno{(3.12)}
$$
Using that $\text{dist}(\Omega_3(n),\text{supp}\nabla\phi_n)\geq
2\alpha^{-1}\log n$ and $|\nabla\phi_n|, |\Delta \phi_n| \leq
c(\log n)^{-1}$, the estimate $(3.9)$ implies that
$$
\norm{\chi_{\Omega_3(n)}u_{i,n}},\hskip1ex
\norm{\chi_{\Omega_3(n)}\nabla u_{i,n}} \leq C(n^2\log n)^{-1},
\qquad i = 1, \ldots, N(n). \eqno{(3.13)}
$$
We now define $v_{i,n} := (1 - \psi_n) u_{i,n}$ and let $\MM_n :=
\hbox{\rm span} \{v_{i,n} ; i = 1, \ldots, N(n) \}$. We claim that
$$
    \dim \MM_n = N(n), \qquad n \ge n_0,
   \eqno{(3.14)}
$$
for some $n_0 \in \N$. Let $H_{Q_n \setminus \Omega_4(n)}$ be the
operator $-\Delta + W$ on $Q_n\backslash\Omega_4(n)$ with
Dirichlet boundary conditions. The functions
$v_{i,n}:=(1-\psi_n)u_{i,n}$ are approximate eigenfunctions of
$H_{Q_n \setminus \Omega_4(n)}$: in fact, using (3.13), one easily
checks that
$$
   \norm{(H_{Q_n \setminus \Omega_4(n)}-E_{i,n})v_{i,n}}\leq C (n^2 \log^2n)^{-1}
 \eqno{(3.15)}
$$
and
$$
  \norm{v_{i,n}-u_{i,n}}\leq C(n^2\log n)^{-1},
       \eqno{(3.16)}
$$
for $i = 1, \ldots, N(n)$. Now (3.7) and (3.16) imply
$\sum_{i=1}^{N(n)} \norm{u_{i,n} - v_{i,n}}^2 < 1$ for $n$ large
and we obtain $(3.14)$.

(3) We next show that there is $n_1 \ge n_0 \in \N$ such that
$$
  \scapro{H_{Q_n \setminus \Omega_4(n)}w}{w} < b \norm{w}^2, \qquad w \in \MM_n, \quad n \ge n_1.
  \eqno{(3.17)}
$$
For a proof, consider an arbitrary
$w=\sum_{i=1}^{N(n)}\gamma_iv_{i,n}\in \MM_n$ with $\norm{w}=1$.
Here we first observe that the coefficients $\gamma_i$ satisfy a
bound $|\gamma_i| \le 2$, for $n$ large, since (writing $\gamma^2
:= \sum_i |\gamma_i|^2$ and $\eta_n^2 := \sum_i \norm{v_{i,n} -
u_{i,n}}^2$)
 $$
 1 = \norm{w}
    \ge |\!| \sum_{i=1}^{N(n)} \gamma_i u_{i,n} |\!|
    - \sum_{i=1}^{N(n)} |\gamma_i| \cdot  |\!| v_{i,n} - u_{i,n} |\!|
   \ge \gamma ( 1 - \eta_n),
  \eqno{(3.18)}
$$
where $\eta_n \to 0$ as $n \to \infty$ by (3.16). Using (3.16) and
the fact that $\nabla \psi_n$ and $\Delta \psi_n$ have support in
$\Omega_3(n) \setminus \Omega_4(n)$, it follows that for $n$ large
$$
  \norm{w}^2=\sum_{i=1}^{N(n)}|\gamma_i|^2+r,\quad
  \scapro{H_{Q_n \setminus \Omega_4(n)} w}{w}
  =\sum_{i=1}^{N(n)}E_i|\gamma_i|^2+r',
  \eqno{(3.19)}
$$
where $r,r'\leq C (\log n)^{-2}$, so that
$$
      \scapro{H_{Q_n \setminus \Omega_4(n)} w}{w} \leq b'\norm{w}^2+r'',
   \eqno{(3.20)}
$$
with $ r''\leq C (\log n)^{-2}$, for $n$ large, and we obtain
(3.17).

(4) We conclude from (3.17) that $\MM_n \subset P_{(-\infty,
b)}(H_{Q_n \setminus \Omega_4(n)})$
    and then (3.14) implies that  $\dim  P_{(-\infty, b)}(H_{Q_n \setminus
    \Omega_4(n)}) \ge \dim \MM_n = N(n)$.
 On the other hand,  min-max arguments  yield an upper bound
for $\dim  P_{(-\infty, b)}(H_{Q_n \setminus \Omega_4(n)})$ of the
form $c n \log n$, and we are done.
\endproof
\vskip1em
  It seems to be possible, using more powerful methods,
 to remove the logarithmic factor in (3.2) and (3.6) (H.~Cornean, private
communication, 2010).
%
%
%
\section{Muffin tin potentials}
In this section, we present a class of examples where one can
arrive at rather precise statements that illustrate some of the
phenomena described before. Note that the results given below are
derived directly, without recourse to Section 2. We will look at
three types of muffin tin potentials and discuss the effect of the
``filling up'' of the gaps at small angles of rotation. We begin
with muffin tins with walls of infinite height, then approximate
by muffin tin potentials of height $n$, for $n \in \N$ large. By
another approximation step, one may obtain examples with
Lipschitz-continuous potentials. These examples show, among other
things, that Schr\"odinger operators of the form $R_\theta$ may in
fact have spectral gaps for some $\theta > 0$.

\vskip1em

 We consider the lattice $\Z^2 \subset \R^2$ where we first
 introduce the Laplacian of a periodic  muffin tin with infinitely high
 walls separating the wells: for $0<r<1/2$, we
 let $D_r :=  B_r({\scriptstyle\frac{1}{2}},
   {\scriptstyle\frac{1}{2}})$ denote  the disc of radius $r$ centered at the point
 $({\scriptstyle\frac{1}{2}}, {\scriptstyle\frac{1}{2}}) \in \R^2$, and
 generate from $D_r$ the periodic sets
 $$
     \Omega_r := \cup_{(i,j) \in \Z^2} (D_r + (i,j)), \qquad 0 < r < 1/2.
   \eqno{(4.1)}
 $$
 The Dirichlet Laplacian $H_r$ of $\Omega_r$ is the direct sum of
 a countable number of copies of the Dirichlet Laplacian on $D_r$;
 therefore, the spectrum of $H_r$ consists in a sequence of
 positive eigenvalues $(\mu_k(r))_{k\in\N}$ with $\mu_k(r) \to \infty$ as
 $k\to\infty$; we may assume that $\mu_k(r) < \mu_{k+1}(r)$ for all $k \in
 \N$.
 The eigenvalues $\mu_k = \mu_k(r)$ of $H_r$ have infinite multiplicity.
 The $\mu_k$ correspond to the bands of a periodic
 problem: in fact, defining $V_r \colon \R^2 \to \R$ by
$$
   V_r(x,y) := \left\{%
\begin{array}{ccc}
  0, & & \text{$(x,y) \in \Omega_r$}, \\
  1, & & \text{$(x,y) \notin \Omega_r$}, \\
\end{array}%
\right. \eqno{(4.2)}
$$
 the periodic Schr\"odinger operators $H_{r,n} := H_0 + nV_r$
 have  purely a.c.\ spectrum with a band/gap structure. Furthermore,
 norm resolvent convergence $H_{r,n} \to H_r$, obtained as in
 [HH], implies that  the bands  of $H_{r,n}$ converge to the eigenvalues
 $\mu_k$ of $H_r$. In the sequel, denote by $(a,b)$ one of the gaps
 $(\mu_k, \mu_{k+1})$.
We next look at the rotation problem where we define
 $$
     \Omega_{r,\theta} := \left(\Omega_r \cap \{ x \ge 0\}\right)
       \cup \left((M_\theta
     \Omega_r) \cap \{x < 0 \} \right);
   \eqno{(4.3)}
 $$
we also let $H_{r,\theta}$ denote the Dirichlet Laplacian on
$\Omega_{r,\theta}$, for $0<r<1/2$ and $0\le\theta\le \pi/4$.

 The set $(M_\theta \Omega_r) \cap \{ x < 0 \}$ comes with two types of connected
 components: most (or, in some cases, all) components are discs, but
 typically there are also discs in $M_\theta \Omega_r$ with center at a distance
 less than $r$ from  the $y$-axis; those appear in  $(M_\theta \Omega_r) \cap \{
 x < 0 \}$ in a truncated form. It is then clear that $H_{r,\theta}$ has
 pure point spectrum.

 Let us comment on some special cases before we proceed:
 for $\tan\theta$ rational, these truncated discs form a periodic
 pattern; furthermore, we will find a half-disc in $(M_\theta \Omega_r) \cap \{ x < 0 \}$
 if and only if there is a disc  in $M_\theta \Omega_r$ with center on the
 $y$-axis  which happens if and only if
 $\tan\theta = 1/(2k+1)$ for some $k \in \N$. It follows that for
 any $\tan\theta \in \Q$ with $\tan\theta \notin \{ 1/(2k+1) \: ; \: k \in \N\}$
 there is some $r_0 > 0$ such that {\it no} component of $M_\theta \Omega_r$ meets the
 $y$-axis, for $0 < r < r_0$; in other words, in this case all components of
 $ \Omega_{r,\theta}$ are discs.

\vskip1em
\centerline{\includegraphics[width=5cm]{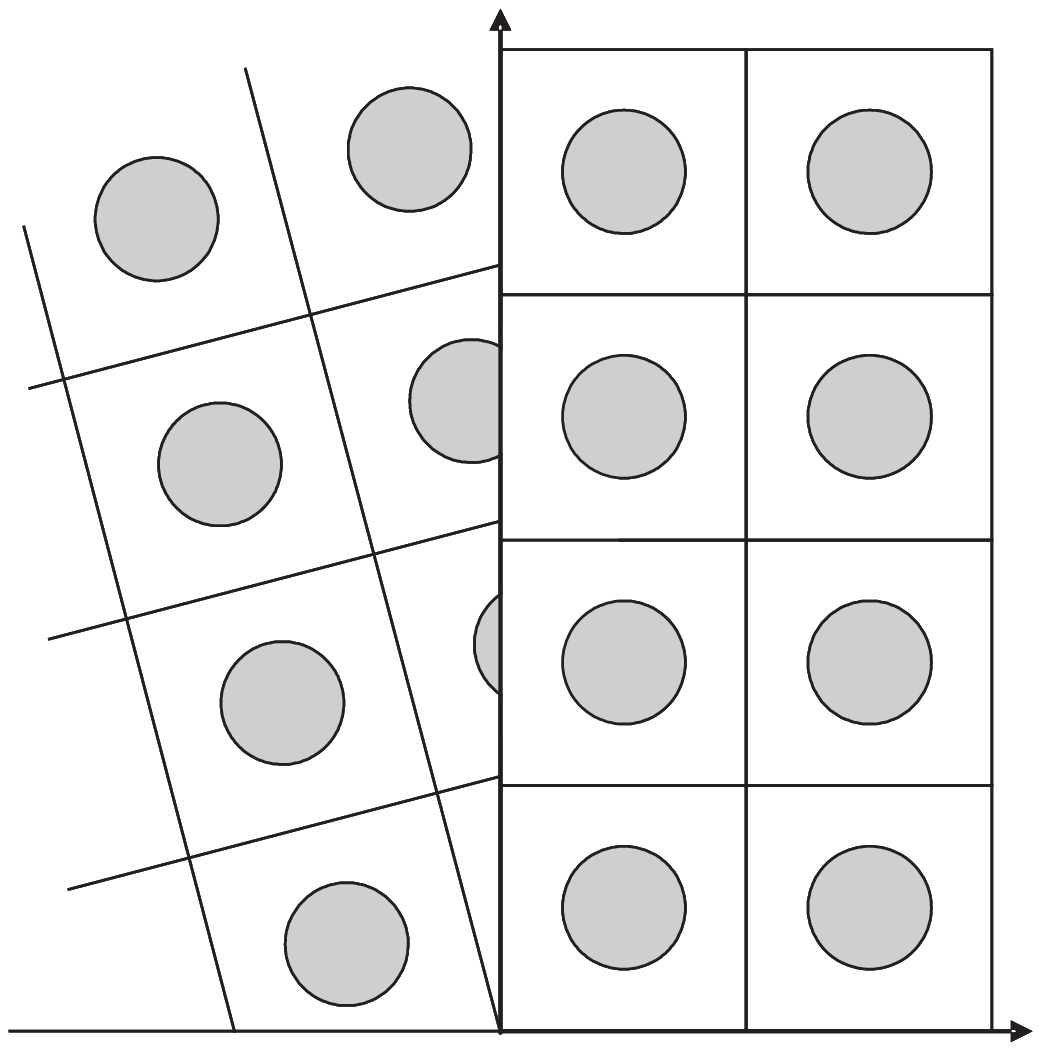}}
\vskip1ex \centerline{Figure 2: The domain $\Omega_{1/4, \pi/8}$
(shaded).} \vskip1em

 We now return to the general situation. Moving up the $y$-axis from the origin, we
 denote the discs in $M_\theta \Omega_r$ that intersect the $y$-axis
 by $D_{j;r,\theta}$, $j \in J_\theta$, with centers $(\xi_{j;\theta}, \eta_{j;\theta})$;
 here $J_\theta = \emptyset$ or $J_\theta = \N$ according to the cases
 discussed above. Without restriction, we may assume that the $\eta_{j;\theta}$ are
  monotonically increasing. Let
  $$
        C_{j ; r,\theta} :=  D_{j ; r,\theta} \cap \{ x < 0 \}, \qquad j \in
        J_\theta.
   \eqno{(4.4)}
 $$
 Clearly, the Dirichlet eigenvalues of the components
 $C_{j;r,\theta}$ are the surface states in this model.
 Since the eigenvalues of the $C_{j;r,\theta}$ only depend on $r$ and $\xi_{j;\theta}$
 (but not on $\eta_{j;\theta}$), it is enough to study the eigenvalues of the following
 sample domains:
 For $r > 0$ and $-r < \xi \le r$, we denote 
 $$
   C_{r,\xi} := \{ (x,y) \in D_r \: ; \: x < 1/2 + \xi \}.
     \eqno{(4.5)}
 $$
 The Dirichlet Laplacian of $C_{r,\xi}$ has eigenvalues $\la_k(r,\xi)$, $k \in
 \N$, which depend continuously on $\xi$, for $k \in \N$ and $r > 0$ fixed.
 Furthermore, $\la_k(r,\xi) \to \infty$ as $\xi \to -r$
 and $\la_k(r,\xi) \to \mu_k(r)$, as $\xi \to r$. Therefore, for each
 $k \in \N$, the eigenvalues  $\la_k(r,\xi)$ range over the interval $[\mu_k(r), \infty)$.

 We now combine the above properties of $\la_k(r,\xi)$
 with information on the distribution of the coordinates $\xi_{j;\theta}$.
 It is easy to see that, as $0 \ne \theta \to 0$, the $\xi_{j;\theta}$ partition
 the interval $(-r,r)$ into subintervals of smaller and smaller length.
 Therefore, for $\eps > 0$ given, any interval
 $(\alpha,\beta) \subset (a,b)$ of length $\ge \eps$ will contain an
 eigenvalue of $H_{r,\theta}$ for all sufficiently small $\theta > 0$.

 We expect stronger properties for angles $\theta$
 for which the set  $\{\xi_{j;\theta} \: ; \: j \in \N\}$
 is dense in $(-r,r)$.  As in Section 2, ergodic theory gives us a set
 $\Theta \subset (0,\pi/2)$ of full measure such that for each
 $\theta \in \Theta$ the set  $\{ (m \tan\theta)_\sim \: ; \:
 m \in \N\}$  is dense in $(0,1)$, which  implies the
 desired property for the $\xi_{j;\theta}$. It follows that, for any $\theta \in
 \Theta$, the eigenvalues of $H_{r,\theta}$ will be dense in $[\mu_1(r),\infty)$.

 Finally, for $\tan\theta$ rational the $\xi_{j;\theta}$  form a
 periodic set and then $H_{r,\theta}$ will only have a finite number of
 eigenvalues (each of infinite multiplicity) in the gap $(a,b)$.
 We thus have the following result:
\vskip1em
\noindent
{\bf 4.1.~Proposition.} {\it Let $0 < r < 1/2$
be fixed.

$(a)$ Each $\mu_k(r)$, $k = 1, 2, \ldots$, is an eigenvalue of
infinite multiplicity
 of $H_{r,\theta}$, for all $0 \le \theta \le\pi/4$. The spectrum of
 $H_{r,\theta}$ is pure point, for all $0 \le \theta \le \pi/4$.

$(b)$ For any $\eps > 0$ there is a $\theta_\eps = \theta_\eps(r)
> 0$ such that any interval
 $(\alpha, \beta) \subset (a,b)$ with $\beta - \alpha \ge \eps$ contains an eigenvalue of
 $H_{r,\theta}$ for any $0 < \theta < \theta_\eps$.

$(c)$ There exists a set $\Theta \subset (0,\pi/2)$ of full
measure such that $\sigma(H_{r,\theta}) = [\mu_1(r),\infty)$.
 The eigenvalues different from the $\mu_k(r)$ are of finite multiplicity.
}
 \vskip1em

\noindent
 {\bf 4.2.~Remark.} Let $\Lambda := \{ \theta \in (0,\pi/2) \: ; \: \tan
\theta \in \Q \}$ denote the set of angles where $\tan\theta$ is rational;
clearly, $\Theta \cap \Lambda = \emptyset$.
It is easy to see that $H_{r,\theta}$, for $\theta \in \Lambda$, has at most a finite number
 of eigenvalues in $(a,b)$, each of them of infinite multiplicity.
 Hence we see a  drastic change in the spectrum for $\theta \in \Lambda$ as
 compared with $\theta \in \Theta$.
 Furthermore, if $\theta \in \Lambda$ with $\tan\theta \notin \{ 1/(2k+1) \: ; \: k \in
 \N \}$, then there is some $r_\theta > 0$ such that $\sigma(H_{r,\theta}) =
 \sigma(H_r)$ for all $0 < r < r_\theta$.

\vskip1em

We next turn to muffin tin potentials of finite height. Here we
define the potential $V_{r,\theta}$ to be zero on
$\Omega_{r,\theta}$ and $V_{r,\theta} = 1$ on the complement of
$\Omega_{r,\theta}$, where $0 < r < 1/2$ and $0 \le \theta \le
\pi/4$; we also let
 $H_{r,n,\theta} := H_0 + n V_{r,\theta}$.  The periodic operators $H_{r,n,0}$
 have purely absolutely continuous spectrum.

 We first show that, for
 $r, \theta$ fixed, the operators $H_{r,n,\theta}$ converge to
 $H_{r,\theta}$ in the sense of norm resolvent convergence.
 This can be seen as follows:
In view of Theorem A.1, we introduce an additional Dirichlet
 boundary condition on a (closed) set
 $S = S_\theta \subset \R^2 \setminus \Omega_{r,\theta}$, which we now
 define:  Let $\Gamma \subset \R^2$ denote
the grid $\{(x,y) \in \R^2 \: ; \: x \in \Z \hbox{\rm\ or\ } y \in
\Z\}$, and let $\Gamma_\theta := M_\theta \Gamma$ denote the
rotated grid. For $\rho := (1/2 -r)/2$, we let $S$ consist of
  $\Gamma \cap \{ (x,y) \: ; \: x > \rho \}$, $\Gamma_\theta \cap \{x < \rho \}$ plus
 the vertical line $\{ x = \rho \}$.
 Note that $S$ has distance $\rho > 0$ from $\Omega_{r,\theta}$.
 Let $H_{r,n,\theta;S}$ denote the operator $-\Delta + n V_{r,\theta}$
 with Dirichlet boundary condition on $S$. We now have
\begin{align}
\lnorm{(H_{r,\theta} + 1)^{-1} \oplus 0 - (H_{r,n,\theta} +
1)^{-1}}
& \le \lnorm{(H_{r,\theta} + 1)^{-1} \oplus 0 -
(H_{r,n,\theta;S} + 1)^{-1}} \nonumber\\
& \quad\quad\quad + \lnorm{ (H_{r,n,\theta;S} + 1)^{-1} -
(H_{r,n,\theta} +
  1)^{-1}}
  \nonumber\\&\hspace{6.4cm}{(4.6)}\nonumber
\end{align}
\vskip-.25cm
\noindent where $0$ denotes the zero operator on $L_2(\R^2
\setminus \Omega_{r,\theta})$. By Theorem A.1, applied with
$U := \R^2 \setminus \overline{\Omega}_{r,\theta}$, we can find $n_0
\in \N$ such that the second term on the RHS is smaller than any
given $\eps>0$, for $n \ge n_0$. The first term on the RHS is a
direct sum of operators living on the components of $\R^2
\setminus S_\theta$. By routine arguments, we have norm resolvent
convergence on each of the components, as $n \to \infty$, and all
we have to do is to convince ourselves that this convergence holds
uniformly on all components.
 The components of $\R^2 \setminus S_\theta$ fall into 4 classes:
there are unrotated and rotated squares, there are rectangles in
the right half-plane of the form $(\rho,1)\times (\ell, \ell +1)$
for $\ell \in \Z$, and there are polygons (triangles, quadrangles,
and pentagons) in the set $\{x < \rho\}$ that are bounded to the
right by the line $x = \rho$. We have no problem with uniform
convergence for the first three classes and Lemma A.2 takes care
of the fourth class. It is easy to see that the norm resolvent
convergence $H_{r,n,\theta} \to H_{r,\theta}$ as $n\to\infty$ is
uniform in $\theta \in [0,\pi/4]$. We then obtain from Proposition
4.1 (b) and Remark 4.2 the following results:

\vskip1.5em

\noindent{\bf 4.3.~Proposition.} {\it

$(a)$ For $\tan\theta \in \Q$ the spectrum of $H_{r,n,\theta}$ has
gaps inside the interval $(a,b)$ for $n$ large. More precisely, if
$H_{r,\theta}$ has a gap $(a',b') \subset (a,b)$, then, for $\eps
> 0$ given,  the interval $(a'+\eps, b'-\eps)$ will be free of
spectrum of $H_{r,n,\theta}$ for $n$ large.

$(b)$ For any $\eps > 0$ there are $\theta_0 > 0$ and $n_0 > 0$
 such that any interval $(c-\eps, c+\eps) \subset (a,b)$ contains
 spectrum of $H_{r,n,\theta}$ for all $0<\theta<\theta_0$ and $n \ge n_0$.

} \vskip1em

By similar arguments, we can approximate $V_{r,\theta}$ by
Lip\-schitz-continuous muffin tin potentials that converge
monotonically (from below) to $V_{r,\theta}$ in such a way that
 norm resolvent convergence holds for the associated
Schr\"odinger operators (again uniformly in $\theta \in
[0,\pi/4]$). The spectral properties obtained are analogous to the
ones stated in Proposition 4.3. Note, however, that the statement
corresponding to part $(b)$ in Proposition 4.3 is weaker than the
result of our main Theorem 0.1.

A brief study of translational dislocation problems for muffin tin
potentials can be found in [HK].


\section{Some extensions and remarks}
(1) A simple variant of the rotation problem consists in rotations
in the left and the right half planes through angles $\theta/2$
and $-\theta/2$, respectively, i.e., we study
$$
{\tilde V}_\theta(x,y) = \left\{%
\begin{array}{lll}
  (V\circ M_{-\theta/2})(x,y), && \text{$x \ge 0$,} \\
  (V \circ M_{\theta/2})(x,y), && \text{$x < 0$;} \\
\end{array}%
\right.\eqno{(5.1)}
$$
this potential might be rather close to the physical situation
shown in Figure~1. Here we consider the accompanying translational
dislocation potentials
$$
{\tilde W}_t(x,y) =\left\{
\begin{array}{lll}
V(x-t/2,y), && \text{$x \ge 0$,} \\
V(x+t/2,y), && \text{$x < 0$.}\\
\end{array}\right.
\eqno{(5.2)}
$$
We may then obtain results as in Theorem 0.1 without the use of
Birkhoff's theorem: here we only need to take care of the second
condition in eqn.~(2.11) since the horizontal alignment between
the left- and right-hand part of $V_\theta$ on the $y$-axis is
guaranteed by the definition of ${\tilde V}_\theta$.

\vskip1ex

(2) We have shown that the spectral gaps of $H$ fill with spectrum
of $R_{\theta}$ as $\theta\to 0$ in the sense that any interval of
length $\eps > 0$ inside a gap of $H$ will contain spectrum of
$R_\theta$ for sufficiently small angles. In general, we do not
know whether the spectrum of $R_\theta$  in the gaps of $H_0$ is
pure point, absolutely continuous or singular continuous.
 However, there are some special angles where we can exclude singular continuous
spectrum: if we assume that $\cos\theta$ is a
 rational number, $\cos\theta=q/p$ with $p,q\in\N$, and $p$ and $q$
belong to a Pythagorean triple ($p^2-q^2 = r^2$ for some
 $r \in \N$), then $V_{\theta}$ has period $p$ in $y$-direction. In
this case, a result in [DS] implies that $\sigma(H_{\theta})$
 has no singular continuous part.
\vskip1ex

(3) It is natural to ask about higher dimensions. Suppose we are
given a potential $V \colon \R^3 \to \R$, periodic with respect to
the lattice $\Z^3$. We may then simply consider rotations of the
$(x,y)$-plane by an angle $\theta$, i.e., we let $V_\theta(x,y,z)
= V(x,y,z)$ in $\{(x,y,z) \: ; \: x \ge 0\}$ and  $V_\theta(x,y,z) =
V(M_{-\theta}(x,y),z)$ in $\{(x,y,z) \: ; \: x < 0\}$, in which case our
methods should apply. However, in $\R^3$ there are many other
rotations for which our methods may or may not work. \vskip1ex

(4) Of course, taking the limit $\theta \to 0$ is a mathematical
idealization. In real crystals or alloys the lattice and its
rotated version have to match up according to certain rules. This
is usually only possible for a small number of angles. Related
questions in higher dimensions are studied under the name of {\it
coincidence site lattices} (CSL); cf.~[B, Z].
\section{Appendix}
In this brief appendix we prove a---rather general---result on
decoupling by Dirichlet boundary conditions placed on a set inside
a high barrier. The method of proof is fairly standard but we have
been unable to find a suitable reference which would have covered
our situation. Let $d \in \N$. For some open set ${U} \subset
\R^d$ and a closed set $S \subset {U}$ of measure zero, we
consider for $n \in \N$ the Schr\"odinger operators $H_n :=
-\Delta +  n \chi_{U}$, acting in $L_2(\R^d)$, and $H_{n,S} :=
-\Delta + n \chi_{U}$ in $L_2(\R^d \setminus S) = L_2(\R^d)$,
where $H_{n,S}$ is assumed to obey Dirichlet boundary conditions
on the set $S$. In other words, the associated quadratic forms
 $h_n$ and $h_{n,S}$ have $\Cci{\R^d}$ and $\Cci{\R^d \setminus S}$ as form
cores. We then show that the resolvent difference $(H_n + 1)^{-1}
- (H_{n,S}+1)^{-1}$ goes to zero in norm, as $n \to \infty$,
provided the set $S$ has a positive distance to the boundary of
${U}$; note that the set $S$ need not be bounded.

\vskip1em\noindent {\bf A.1.~Theorem.} {\it Let $H_n$ and
$H_{n,S}$ as above and suppose that $\dist(S, \partial {U}) >
0$. Then
$$
   \lnorm{(H_n + 1)^{-1} - (H_{n,S}+1)^{-1}} \to 0, \quad n \to \infty.
   \eqno{(A.1)}
$$
}\vskip1em\noindent {\bf Proof.} For $f \in L_2(\R^d)$ with
$\lnorm{f} \le 1$, let $u_n := (H_n + 1)^{-1} f$ and $v_n :=
(H_{n,S} + 1)^{-1} f$. We then have the trivial estimates
$\lnorm{u_n} \le 1$ and
$$
  \lnorm{\nabla u_n}^2 + n \int_{U} |u_n|^2 dx = h_n[u_n]
       \le \scapro{(H_n + 1) u_n}{ u_n} = \scapro{f}{u_n} \le 1,
  \eqno{(A.2)}
$$
so that $\lnorm{\nabla u_n} \le 1$ and $\int_{U} |u_n|^2 dx \le
1/n$, for all $n \in \N$; analogous estimates hold for $v_n$.
We let $\FF$ denote the set of all functions $\phi \in
C^\infty(\R^d)$ that satisfy $\supp\phi \subset {U}$,
$\dist(\supp\phi, \partial{U}) > 0$, $ 0\le \phi \le 1$, and
$\nabla\phi$, $\Delta\phi$ bounded. For any $\phi \in \FF$, we
then have $\phi u_n \in D(H_n)$ and $(H_n + 1) (\phi u_n) = \phi f
- 2 \nabla u_n \nabla \phi - u_n \Delta \phi$. We immediately see
from the above estimates that there is a constant $c_\phi \ge 0$
such that
$$
   \lnorm{H_n (\phi u_n)} \le c_\phi.
   \eqno{(A.3)}
$$
We now derive a crucial estimate for $\lnorm{\nabla (\phi u_n)}$:
Since $\lnorm{\nabla (\phi u_n)}^2 \le h_n[\phi u_n]
  = \scapro{H_n (\phi u_n)}{ \phi u_n}$, for any $\eps > 0$ there are constants
 $C_\eps, C_\eps' \ge 0$ such that
$$
  \lnorm{\nabla (\phi u_n)}^2
           \le \eps \lnorm{H_n (\phi u_n)}^2 + C_\eps \lnorm{\phi u_n }^2
            \le \eps c_\phi^2 + C_\eps'/ n.
    \eqno{(A.4)}
$$
Analogous estimates hold for $H_{n,S}$ and $v_n$.

 We let $\rho := \dist(S, \partial{U})$, $U_\rho :=
 \{ x \in {U} {\: ; \:} \dist(x,S) < \rho/2\}$ and fix some function $\psi \in \FF$
 with $\psi = 1$ on $U_\rho$. Let $\eta := 1 - \psi$ and choose
 another function $\phi_\psi \in \FF$ satisfying
 $\phi_\psi = 1$ on the support of $\nabla\psi$.
 Writing $w_n = u_n - v_n = (H_n + 1)^{-1} f - (H_{n,S} + 1)^{-1} f$ it is then
 clear that $\psi w_n \to 0$, as $n \to \infty$, uniformly for all $f$
 with $\norm{f} \le 1$, and it remains to consider $\eta w_n$.
 Since $\eta$ vanishes on $U_\rho$, we have $\eta w_n \in D(H_n)$ and
$$
  (H_n + 1) (\eta w_n) = - 2 \nabla \eta \cdot \nabla w_n - (\Delta \eta) w_n =:
  z_n,
  \eqno{(A.5)}
$$
so that, in particular,
$$
   \eta w_n = (H_n + 1)^{-1} z_n, \qquad \lnorm{\eta w_n} \le \lnorm{z_n}.
   \eqno{(A.6)}
$$
We finally show that $z_n \to 0$, uniformly for $\lnorm{f} \le 1$.
Here we know already that $(\Delta \eta) w_n \to 0$ since $\Delta
\eta$ is supported inside ${U}$. Applying the estimate (A.4) we
find that
$$
  \lnorm{\nabla \eta \cdot \nabla w_n} \le C_\eta \lnorm{\nabla (\phi_\psi w_n)}
   \le \eps c_\psi + C'_{\eps,\psi}/n,
  \eqno{(A.7)}
$$
with $c_\psi$ as in (A.3), (A.4). For $\delta > 0$ given, we can
find $\eps > 0$ s.th.\ $\eps c_\psi < \delta/2$ and then $n_0 \in
\N$ s.th.\ $C'_{\eps,\psi} /n < \delta/2$ for all $n \ge n_0$, and
we are done.
\endproof
\vskip1em\noindent {\bf Remarks.}
(a) The assumption that $S$ has measure zero is made chiefly for
simplicity of notation.

(b) The same result holds for Schr\"odinger operators $-\Delta + n
V$ where $V \colon \R^d \to \R$ is a non-negative, bounded
potential satisfying $V \ge \chi_{U}$. Our method of proof
yields a bound on the norm of the resolvent difference which is
independent of $V$.

(c) It is well-known that $H_n$ and $H_{n,S}$ converge to a
suitably defined Dirichlet Laplacians in the sense of strong
resolvent convergence, cf.~ [HZh]. For periodic ${U}$ the
convergence would even be in the norm resolvent sense, cf.~[HH].

\vskip1em

We finally discuss uniform norm resolvent convergence for
Schr\"odinger operators on domains of the type encountered at the
end of Section 4. Here we will use the well-known fact that a
monotonic sequence $(A_n)_{n\in\N}$ of compact operators which
converges strongly to a (compact) operator $A$ converges in norm,
i.e., $\norm{A_n - A} \to 0$.

For simplicity of notation, we consider an equivalent geometric
situation with a family of domains in the first quadrant given as
follows:

We fix $0 < \theta < \pi/4$ and $0 < r, d < 1/2$. Then, for $s \in
\R$,  we consider two parallel lines $\ell_{1,s}$, $\ell_{2,s}$
defined, respectively, by the equations $y = (x - s) \cot \theta$
 and $y = (x - s) \cot\theta + d/\sin\theta$ so that
$\ell_{1,s}$ and $\ell_{2,s}$ have distance $d$. We let
$$
  G_s := [0,1]^2 \cap \{ (x,y) {\: ; \:} y > \ell_{1,s}(x) \},
  \qquad D_{r,s} := B_r({\scriptstyle\frac{1}{2}},
   {\scriptstyle\frac{1}{2}}) \cap  \{ (x,y) {\: ; \:} y > \ell_{2,s}(x) \};
 \eqno{(A.8)}
$$
cf.\ Figure 3. Note that $D_{r,s}$ (if it is non-empty) has
distance $\min\{d, 1 - r\}$ to the boundary of $G_s$.

\vskip1em
\centerline{\includegraphics[width=5cm]{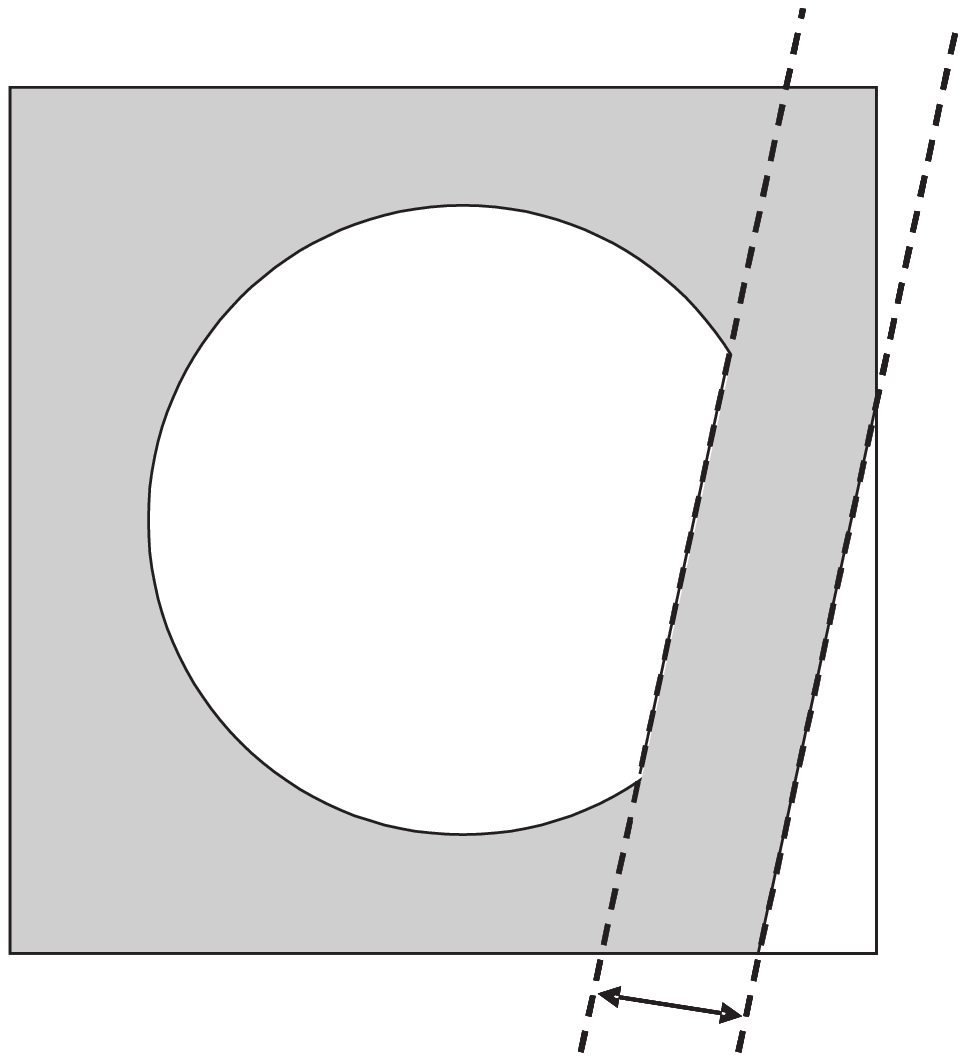}}\vskip1em
\centerline{Figure 3: A typical domain $D_{r,s}$ (shaded).}
\begin{picture}(1,1)
\put(157,108){$D_{r,s}$}%
\put(110,60){$G_s$}%
\put(190,30){$d$}%
\put(198,190){$\ell_{2,s}$}%
\put(223,190){$\ell_{1,s}$}%
\end{picture}
\vspace{.0cm}
%
%

We next define quadratic forms on the domains $G_s$:
 for $n \in \N$, we let $V_{r,s,n} := 0$ on $D_{r,s}$ and
 $V_{r,s,n} := n$ on $G_s \setminus D_{r,s}$. As $s$ increases,
the sets $G_s$ and $D_{r,s}$ both increase and we therefore see
that the quadratic forms
$$
   h_{r,s,n}[\phi] := \norm{\nabla \phi}^2 + \int_{G_s} V_n |\phi|^2 \d x,
   \qquad \phi \in \Cci{G_s},
 \eqno{(A.9)}
$$
depend monotonically on $s$. It is easy to include the case $n =
\infty$ by setting
$$
     h_{r,s,\infty}[\phi] := \norm{\nabla \phi}^2, \qquad \phi \in
     \Cci{D_{r,s}}.
  \eqno{(A.10)}
$$
The self-adjoint operators $H_{r,s,n}$ and $H_{r,s,\infty}$
associated with (the closure of) these quadratic forms have
compact resolvent. We have the following result:
\vskip1em\noindent {\bf A.2.~Lemma.} {\it With the above
definitions and assumptions, we have }
$$
  \sup_s  \norm{ H_{r,s,n}^{-1} - H_{r,s,\infty}^{-1}} \to 0, \qquad n \to
  \infty.
  \eqno{(A.11)}
$$
\vskip1em\noindent {\bf Proof.}  Obviously, we  may restrict our
attention to parameters $s$ from a compact interval $J  \subset
\R$. Writing
$$
  f_n(s) := \norm{H_{r,s,n}^{-1} - H_{r,s,\infty}^{-1}}, \qquad s \in J,
  \eqno{(A.12)}
$$
monotonicity and compactness imply that the functions $f_n$ are
continuous with $f_n(s) \to 0$  monotonically as $n \to \infty$,
for each fixed $s$. Now the desired result follows by Dini's
Theorem. \endproof

\end{document}